\documentclass[fleqn,twoside]{article}
\usepackage{espcrc2}

\usepackage{graphicx}
\newcommand{\AmS}{{\protect\the\textfont2
  A\kern-.1667em\lower.5ex\hbox{M}\kern-.125emS}}

\newcommand{\etal}{{\it et al.}}

\title{Charm Leptonic and Semileptonic Decays}

\author{P. Zweber\address{University of Minnesota, Minneapolis, Minnesota 55455 USA}}
       
\begin{document}

\begin{abstract}
Experimental results for the pseudoscalar decay constants $f_{D}$ and $f_{D_s}$ are reviewed.  
Semileptonic form factor results from $D \to {\rm (pseudoscalar)}l \nu$ 
and $D \to {\rm (vector)}l \nu$ decays are also reviewed.  
\end{abstract}

\maketitle

\section{INTRODUCTION}

The study of charmed meson decays are important for 
improving our knowledge of the Standard Model (SM).  
In particular, the study of leptonic and semileptonic decays allow us to measure 
the CKM matrix elements $V_{cs}$ and $V_{cd}$ to a high level of precision.  
Leptonic and semileptonic decays are also used to test theoretical predictions 
describing the strong interaction (QCD) in heavy quark systems.  
Charmed meson decays are a good laboratory to test Lattice QCD (LQCD) predictions, 
which can then be applied with confidence to bottom meson decays.  
Improved theoretical predictions will not only lower the uncertainty in the CKM matrix elements 
mentioned above but also $V_{cb}$ and $V_{ub}$.

\section{LEPTONIC DECAYS}

Leptonic decays of heavy mesons involve the annihilation of the constituent quarks 
into a neutrino and charged lepton via a virtual $W$ boson.  
The probability for the annihilation is proportional to the wavefunctions of the quarks 
at the point of annihilation and is incorporated within the decay constant of the meson. 
The leptonic decay partial width of the $D^+_{(s)}$ meson within the SM is given by \cite{LepDecayPW}
\begin{eqnarray}
\nonumber \Gamma(D^+_{(s)} \to l^+ \nu) = \\
 & \hspace*{-75pt}\frac{G^2_F}{8\pi} f^2_{D_{(s)}} m^2_l M_{D_{(s)}} 
\left(1-\frac{m^2_l}{M^2_{D_{(s)}}}\right)|V_{cd(s)}|^2,
\label{eq:LepDecPartWidth}
\end{eqnarray}
where $G_F$ is the Fermi coupling constant, 
$f_{D}$ ($f_{D_s}$) is the $D^+$ ($D^+_s$) decay constant, 
$M_{D}$ ($M_{D_s}$) is the $D^+$ ($D^+_s$) mass, 
$m_l$ is the mass of the charged lepton, 
and $V_{cd}$ ($V_{cs}$) is the $c\to d$ ($c\to s$) CKM transition matrix element.  
These decays are helicity suppressed and, based on Eq. \ref{eq:LepDecPartWidth}, 
the SM predicts 
$\Gamma(D^+ \to l^+ \nu) = 2.3\times10^{-5}:1:2.64$ and 
$\Gamma(D^+_s \to l^+ \nu) = 2.3\times10^{-5}:1:9.72$, with $l = e:\mu:\tau$, respectively.  
Deviations from these predictions may arise from physics beyond the SM.  
Recent measurements of leptonic decays of the $D^+$ and $D^+_s$ are discussed 
in the following subsections.

\subsection{$D^+ \to l^+ \nu$}

The BES Collaboration reported a measurement of the 
$D^+ \to \mu^+ \nu_{\mu}$ branching fraction using 33 pb$^{-1}$ 
of $e^+ e^-$ annihilation data collected on the $\psi(3770)$ resonance 
with the BES II detector \cite{BESfD}.  
Events are selected using the Mark III ``$D$-tagging'' method \cite{MarkIIITag},
which consists of fully reconstructing the $\overline{D}$ meson from  
$e^+ e^- \to \psi(3770) \to D \overline{D}$ events and studying the $D$ decay.  
In (semi)leptonic decays, the neutrino in inferred from the missing four-momentum in the event.  
Using a sample of 5300 tagged $D^-$ decays (note that charge conjugation is implied unless 
otherwise stated) from 9 tag modes 
and requiring one additional particle consistent with a muon, 
3 candidates events are observed with 0.3 background events.  
This leads to a branching fraction of 
${\cal B}(D^+ \to \mu^+ \nu_{\mu}) = (12^{+11}_{-5}\pm 1)\times10^{-4}$, 
where the first error is statistical and the second is systematic.  

\begin{figure}[htb]
\includegraphics[scale=0.43]{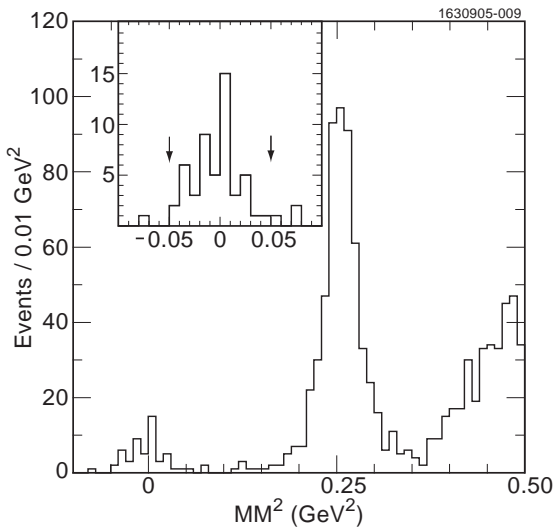}
\caption{CLEO $MM^2$ distribution for $D^+ \to \mu^+ \nu_{\mu}$ decays \cite{CLEOfD}.  
The inset shows the signal region, defined by the arrows, in more detail.  
The background peak at $MM^2 \sim 0.25~{\rm GeV}^2$ is consistent with $D^+ \to K^0_L \pi^+$ decays 
and is substantially displayed from the signal region.}
\label{fig:CLEODtomunu}
\end{figure}

The CLEO Collaboration reported an updated measurement of the 
$D^+ \to \mu^+ \nu_{\mu}$ branching fraction \cite{CLEOfD}.  
Using a 281 pb$^{-1}$ data sample collected at the $\psi(3770)$ resonance with 
the CLEO-c detector,  a sample of 158,000 $D^-$ decays from 6 tag modes was studied using 
the $D$-tagging method described above.  
Candidate events are required to have one charged track of opposite charge to the 
tagged $D^-$ and the track needs to deposit an energy in the electromagnetic calorimeter 
($E_{{\rm tkCC}}$) 
$< 300$ MeV, which is consistent with a minimum ionizing particle.  
Events with an isolated photon-like shower in the calorimeter 
with an energy in excess of 250 MeV are rejected.   
Figure \ref{fig:CLEODtomunu} shows the missing mass squared distribution for the candidate events, 
where the missing mass squared is defined as 
$MM^2 \equiv (E_{{\rm b}} - E_{\mu})^2 - (-{\bf p}_{D} -{\bf p}_{\mu})^2$, 
where $E_{{\rm b}}$ is the beam energy, 
${\bf p}_{D}$ is the momentum of the tagged $D^-$, 
and $E_{\mu}$ ($\bf{p}_{\mu}$) is the energy (momentum) of the muon.  
The signal region, defined by $|MM^2| < 0.05$ GeV$^2$, 
contains 50 candidate and 2.8 background events, of which 1.1 events 
are determined from the SM prediction for 
$D^+ \to \tau^+ \nu_{\tau}, \tau^+ \to \pi^+ \overline{\nu}_{\tau}$. 
The result corresponds to ${\cal B}(D^+ \to \mu^+ \nu_{\mu}) = (4.4\pm0.7\pm0.1)\times10^{-4}$.
  
A complementary CLEO analysis searched for $D^+ \to e^+ \nu_e$ decays \cite{CLEOfD}.  
In this case, the charged track is required to be consistent with a positron.  
No signal is observed and an upper limit of 
${\cal B}(D^+ \to e^+ \nu_e) < 2.4\times10^{-5}$ at 90$\%$ confidence level (C.L.) is obtained.

The CLEO Collaboration has reported on the search for 
$D^+ \to \tau^+ \nu_{\tau}, \tau^+ \to \pi^+ \overline{\nu}_{\tau}$ using the same 281 pb$^{-1}$ 
and tagged $D^-$ sample \cite{CLEODtotaunu}.    
For minimum ionizing pions with $E_{{\rm tkCC}} < 300$ MeV, 
the signal region is defined as 0.05 $< MM^2 <$ 0.175 GeV$^2$, 
above $D^+ \to \mu^+ \nu_{\mu}$ signal region and below the $D^+ \to K^0_L \pi^+$ background 
(see Fig. \ref{fig:CLEODtomunu}).
Pions which hadronize in the calorimeter are studied by requiring the track 
to have $E_{{\rm tkCC}} > 300$ MeV, be inconsistent with a positron, 
and reside in the region -0.05 $< MM^2 <$ 0.175 GeV$^2$.  
Events are rejected in both cases if it contains an isolated photon-like shower 
with an energy in excess of 250 MeV or if the candidate pion is more consistent with being a charged 
kaon.  No significant enhancement is observed in either case, resulting in an upper limit of 
${\cal B}(D^+ \to \tau^+ \nu_{\tau}) < 2.1\times10^{-3}$ (90\% C.L.).
Using the CLEO result for $D^+ \to \mu^+ \nu_{\mu}$ above, 
an upper limit for the ratio ${\cal B}(D^+ \to \tau^+ \nu_{\tau})/{\cal B}(D^+ \to \mu^+ \nu_{\mu})$ 
of 1.8 times that of the SM prediction is placed.

\subsection{$D^+_s \to l^+ \nu$}

\begin{figure}[htb]
\includegraphics[scale=0.37]{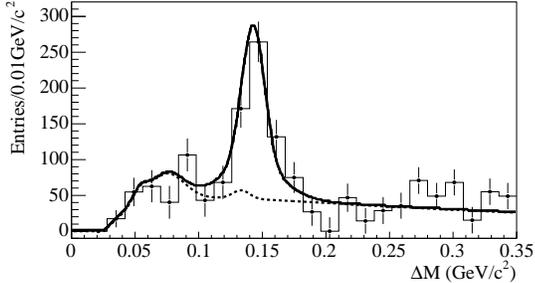}
\caption{BaBar $\Delta M$ distribution for $D^+_s \to \mu^+ \nu_{\mu}$ decays \cite{BABARfDs}.  
The solid line is the fitted signal and background distribution, while the dashed line is the 
background distribution alone.}
\label{fig:BABARDstomunu}
\end{figure}

The BaBar Collaboration analyzed a 230.2 fb$^{-1}$ data sample taken at or near the $\Upsilon(4S)$ 
resonance to study the decay process $D^+_s \to \mu^+ \nu_{\mu}$ \cite{BABARfDs}.  
They studied the process $e^+ e^- \to D^{*+}_s D_{{\rm tag}}$, 
where $D_{{\rm tag}}$ are 16 fully reconstructed $D^0, D^-, D^-_s$, and $D^{*-}$ decays and 
$D^{*+}_s \to \gamma D^+_s \to \gamma (\mu^+ \nu_{\mu})$, 
using a method similar to an earlier CLEO measurement \cite{CLEOfDs98}.  
Figure \ref{fig:BABARDstomunu} shows the resultant $\Delta M \equiv M(D^{*+}_s) - M(D^{+}_s)$ 
distribution. With $489\pm 55$ signal events, 
they determine $\Gamma(D^+_s \to \mu^+ \nu_{\mu})/\Gamma(D^+_s \to \phi \pi^+) = 0.143\pm0.018\pm0.006$.  
Using the recent BaBar result ${\cal B}(D^+_s \to \phi \pi^+) = (4.71\pm0.46)\%$ \cite{BABARphipi},  
they determine a branching fraction of 
${\cal B}(D^+_s \to \mu^+ \nu_{\mu}) = (0.67\pm0.08\pm0.03\pm0.07)\%$, where the last error 
arises from the $D^+_s \to \phi \pi^+$ normalization uncertainty.

\begin{figure}[htb]
%\vspace{-14pt}
\includegraphics[scale=0.38]{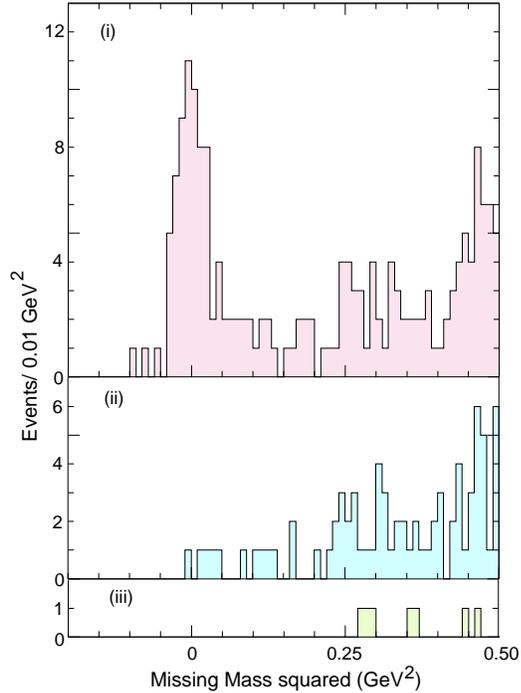}
\caption{CLEO $MM^2$ distributions for $D^+_s \to l^+ \nu$ decays, where $l = e,\mu,\tau$ \cite{CLEOfDs}.  
Figure \ref{fig:Dstomutaunu}i contains tracks with $E_{{\rm tkCC}} < 300$ MeV, 
Fig. \ref{fig:Dstomutaunu}ii contains tracks with $E_{{\rm tkCC}} > 300$ MeV 
and inconsistent with the positron hypothesis, while 
Fig. \ref{fig:Dstomutaunu}iii contains tracks which are consistent with the positron hypothesis.}  
\label{fig:Dstomutaunu}
\end{figure}

As part of the CLEO-c run plan, the energy region 3.97 - 4.26 GeV was scanned to determine 
the best energy for $D_s$ production.  
The energy $\sqrt{s} = 4.170$ GeV was found to be the location with the highest production rate 
of $D_s$ mesons, but the $D_s$ mesons are predominately produced in the processes 
$e^+ e^- \to D^{*-}_s D^+_s , D^-_s D^{*+}_s$.  
The tagging procedure becomes more complicated by the presence of the transition photon 
from the radiative decay process $D^{*}_s \to D_s \gamma$.

Using a 195 pb$^{-1}$ data sample collected near $\sqrt{s} = 4.170$ GeV, 
the CLEO Collaboration reported preliminary branching fractions for 
$D^+_s \to l^+ \nu$ decays, where $l = e,\mu,\tau$, and with the tau decaying 
via $\tau^+ \to \pi^+ {\overline{\nu}_{\tau}}$ \cite{CLEOfDs}.  
Events are selected with 8 $D^-_s$ tag modes and the detection of the transition photon.  
The missing mass squared for $D^+_s$ is defined as 
$MM^2 \equiv (E_{{\rm CM}} -E_{D_s} -E_{\gamma} -E_{tk})^2 
- (-{\bf p}_{D_s} -{\bf p}_{\gamma} -{\bf p}_{{\rm tk}})^2$, 
where $E_{{\rm CM}}$ is the center-of-mass energy, 
$E_{D_s}$ (${\bf p}_{D_s}$) is the energy (momentum) of the tagged $D^-_s$ , 
$E_{\gamma}$ (${\bf p}_{\gamma}$) is the energy (momentum) of the transition photon, and 
$E_{{\rm tk}}$ (${\bf p}_{{\rm tk}}$) is the energy (momentum) of the candidate track.  
The signal side, consisting of one charged track, is reconstructed using the 
same criteria as the CLEO $D^+ \to l^+ \nu$ analyses described above, 
but with the exceptions of increasing the $D^+_s \to e^+ \nu_e$ and  $D^+_s \to \tau^+ \nu_{\tau}$ 
$MM^2$ signal regions to 0.20 GeV$^2$ and increasing the maximum shower energy to 300 MeV.  
Figure~\ref{fig:Dstomutaunu} shows the $MM^2$ distributions. 
Signals are apparent in Figs. \ref{fig:Dstomutaunu}i and \ref{fig:Dstomutaunu}ii, 
while no signal is observed near $MM^2 = 0$ for the positron decay channel in Fig. \ref{fig:Dstomutaunu}iii.  
The $D^+_s \to \mu^+ \nu_{\mu}$ signal region contains 64 candidate and 7.3 background events, 
of which 5.3 events are determined from the SM prediction for 
$D^+_s \to \tau^+ \nu_{\tau}, \tau^+ \to \pi^+ {\overline \nu}_{\tau}$.  
Using events with candidate tracks inconsistent with the positron hypothesis 
and outside of the $D^+_s \to \mu^+ \nu_{\mu}$ signal region, 
36 candidate and 4.8 background $D^+_s \to \tau^+ \nu_{\tau}$ events are found. 
The preliminary branching fractions are 
${\cal B}(D^+_s \to e^+ \nu_e) < 3.1\times10^{-4}$ (90\% C.L.), 
${\cal B}(D^+_s \to \mu^+ \nu_{\mu}) = (0.66\pm0.09\pm0.03)\%$, and 
${\cal B}(D^+_s \to \tau^+ \nu_{\tau}) = (7.1\pm1.4\pm0.3)\%$.  

\begin{table*}[htb]
\caption{Experimental and theoretical results for the decay constants $f_{D}$ and $f_{D_s}$.}
\label{table:decayconstants}
\begin{tabular}{llll}
\hline
   & $f_{D_s}$ (MeV) & $f_{D}$ (MeV) & $f_{D_s}/f_{D}$ \\
\hline
CLEO\cite{CLEOfDs,CLEOStoneICHEP}$^a$\cite{CLEOfD} 
& 280(12)(6)$^a$ & 222.6(16.7)$^{+2.8}_{-3.4}$ & 1.26(11)(3)$^a$ \\
BaBar\cite{BABARfDs} & 283(17)(4)(14) &  & \\
BES\cite{BESfD}  &  & 371$^{+129}_{-117}$(25) &  \\
\hline
Unquenched LQCD\cite{Aubin06} & 249(3)(16) & 201(3)(17) & 1.24(1)(7) \\
Quenched LQCD\cite{Chiu05} & 266(10)(18) & 235(8)(14) & 1.13(3)(5) \\
Quenched LQCD\cite{Lellouch01} & 236(8)$^{+17}_{-14}$ & 210(10)$^{+17}_{-16}$ & 1.13(2)$^{+4}_{-2}$ \\
Quenched LQCD\cite{Becirevic99} & 231(12)$^{+6}_{-1}$ & 211(14)$^{+10}_{-12}$ & 1.10(2) \\
QCD Sum Rules\cite{Gordes05} & 205(22) & 177(21) & 1.16(1)(3) \\
QCD Sum Rules\cite{Narison02} & 235(24) & 203(23) & 1.15(4) \\
Quark Model\cite{Ebert06} & 268 & 234 & 1.15 \\
Quark Model\cite{Cvetic04} & 248(27) & 230(25) & 1.08(1) \\
Potential Model\cite{Salcedo04} & 253 & 241 & 1.05 \\
Isospin Splittings\cite{Amundson93} &  & 262(29) &  \\
\hline
\end{tabular}\\
$^a$The CLEO $f_{D_s}$ result is the average of the 3 preliminary measurements described in the text.
\end{table*}

A complementary CLEO analysis of ${\cal B}(D^+_s \to \tau^+ \nu_{\tau})$ was reported using the 
decay mode $\tau^+ \to e^+ \nu_{e} \overline{\nu}_{\tau}$ \cite{CLEOStoneICHEP}.  
This analysis reconstructs the tagged $D^-_s$ but does not require detection of the transition photon.  
Events are selected with a positron with opposite charge to the tag 
and requires the summed energy of all isolated showers in the calorimeter 
($E^{{\rm extra}}_{{\rm CC}}$) $<$ 400 MeV.  The preliminary branching fraction is  
${\cal B}(D^+_s \to \tau^+ \nu_{\tau}) = (6.3\pm0.8\pm0.5)\%$ 
and, while the CLEO $D^+_s \to \tau^+ \nu_{\tau}, 
\tau^+ \to \pi^+ \overline{\nu}_{\tau}$ analysis has smaller systematic uncertainty,  
this result has a smaller statistical uncertainty.

\begin{figure}[htb]
\includegraphics[scale=0.40]{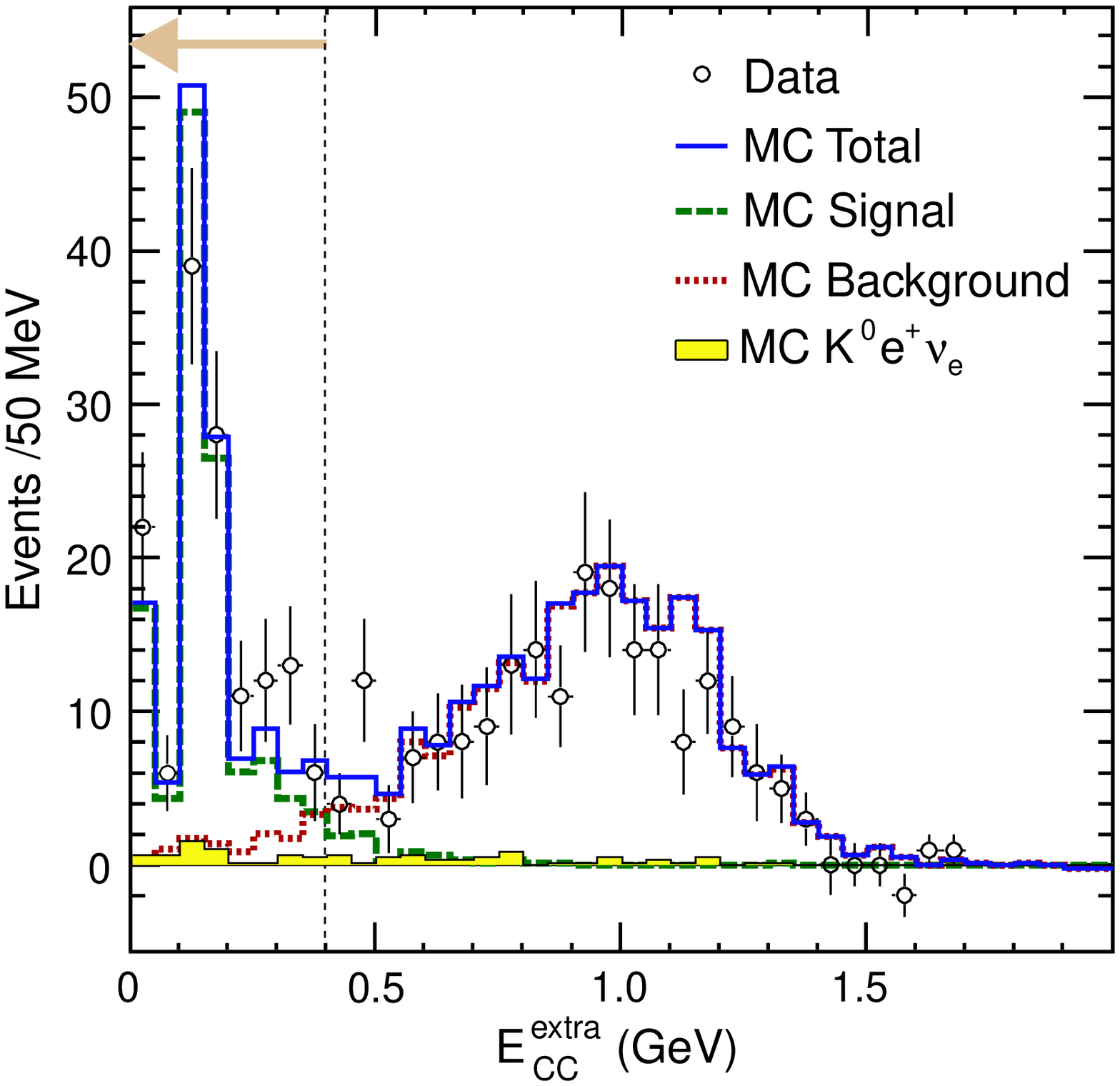}
\caption{$E^{{\rm extra}}_{{\rm CC}}$ energy distribution from the CLEO 
$D^+_s \to \tau^+ \nu_{\tau}, \tau^+ \to e^+ \nu_{e} \overline{\nu}_{\tau}$ analysis \cite{CLEOStoneICHEP}.}  
\label{fig:Dstotauenunu}
\end{figure}

\subsection{Decay Constants $f_{D}$ and $f_{D_s}$}

Table \ref{table:decayconstants} summarizes the experimental results for 
$f_D$ and $f_{D_s}$ along with various theoretical predictions.  
The experimental results are consistent with most theoretical models.  
The decay constant ratio from the CLEO results is $f_{D_s}/f_{D} = 1.26\pm 0.11\pm 0.03$, 
which is consistent with the unquenched LQCD\cite{Aubin06} prediction $1.24\pm 0.01\pm 0.07$ 
and slightly larger than other predictions.

\section{SEMILEPTONIC DECAYS}

\begin{table*}[htb]
\caption{Preliminary CLEO branching fraction measurements of rare semileptonic $D$ decays.  
Results are at the $10^{-4}$ level and upper limits are at 90\% C.L.}
\label{table:CLEOrareD}
\begin{tabular}{lllll}
\hline
 & CLEO \cite{CLEOGaoICHEP} & PDG\cite{PDG06} & ISGW2\cite{ISGW2} & HQS\cite{Fajfer05SL} \\
\hline
${\cal B}(D^+ \to \eta~e^+ \nu_e)$           & 12.9(1.9)(0.7)    & $<$ 70         & 16$^a$  & 10  \\
${\cal B}(D^+ \to \eta^{\prime}~e^+ \nu_e)$  & $<$ 3.0           & $<$ 110        & 3.2$^a$  & 1.6 \\
${\cal B}(D^+ \to \phi~e^+ \nu_e)$           & $<$ 2.0           & $<$ 209        &  &  \\
${\cal B}(D^+ \to \omega~e^+ \nu_e)$           & 14.9(2.7)(0.5)  & 16$^{+7}_{-6}$ & 13  &  \\
${\cal B}(D^0 \to K^-\pi^+\pi^-e^+ \nu_e)$   & 2.9$^{+1.5}_{-1.1}$(0.5)$^{b}$     & $<$ 12 &  &  \\
${\cal B}(D^0 \to K^-_{1}(1270)~e^+ \nu_e,$      &   &                      \\
~~~$K^-_{1}(1270) \to K^-\pi^+\pi^-)^{b}$        & 2.2$^{+1.4}_{-1.0}$(0.2)$^{b}$     &  & 5.6(6)$^c$ &  \\
\hline
\end{tabular}\\
$^a$Prediction assumes 10$^{\circ}$ $\eta-\eta^{\prime}$ mixing angle.  
$^b$Signal observed at a 4.5$\sigma$ significance.\\
$^c$Prediction includes experimental uncertainties from $K^-_{1}(1270) \to K^- \pi^+ \pi^-$ decay\cite{PDG06}.
\end{table*}

\subsection{Inclusive $D$ Decays}

The CLEO Collaboration reported measurements of the inclusive branching fractions for 
$D^0 \to X~e^+ \nu_e$ and $D^+ \to X~e^+ \nu_e$ decays 
and the corresponding positron momentum spectra 
from the 281 pb$^{-1}$ $\psi(3770)$ data sample \cite{CLEOinclD}.  
The positron momentum spectra from the $D^0$ and $D^+$ decays are similar, as expected.  
They determine branching fractions of
${\cal B}(D^0 \to X e^+ \nu_e) = (6.46\pm 0.17\pm 0.13)\%$ and 
${\cal B}(D^+ \to X e^+ \nu_e) = (16.13\pm 0.20\pm 0.33)\%$.  
After subtracting the known exclusive semileptonic decay modes \cite{PDG06}, 
the unobserved semileptonic branching fractions are 
$(0.3\pm 0.3)\%$ and $(1.1\pm 0.7)\%$ for the $D^0$ and $D^+$, respectively.  
It is apparent that the known exclusive decays almost completely saturate the inclusive branching 
fractions. 
Combining these branching fractions with the world average for the $D^{+/0}$ lifetimes \cite{PDG06}, 
the partial width ratio is
$\Gamma(D^+ \to X e^+ \nu_e)/\Gamma(D^0 \to X e^+ \nu_e) = 0.985\pm0.028\pm0.015$, 
consistent with isospin invariance.

\subsection{Rare $D$ Decays}

The CLEO Collaboration has searched for various rare semileptonic $D$ decays 
with its 281 pb$^{-1}$ $\psi(3770)$ sample.  
Table \ref{table:CLEOrareD} lists the preliminary branching fractions.  
With the exception of ${\cal B}(D^+ \to \omega~e^+ \nu_e)$, 
these results are either first observations or improve upper limits by one order of magnitude. 
The results are consistent with predictions based on ISGW2 \cite{ISGW2} 
and heavy quark symmetry (HQS) \cite{Fajfer05SL}.

\subsection{$D \to {\rm (pseudoscalar)}l^+ \nu$}

Precision measurements of the $D^0 \to K^-~e^+ \nu_e$ and $D^0 \to \pi^-~e^+ \nu_e$ branching fractions 
have been measured and are listed in Table \ref{table:DKpienuBF}.  
The BES \cite{BESKpienu} and preliminary CLEO(Tag) \cite{CLEOGaoICHEP} analyses, 
using 33 pb$^{-1}$ and 281 pb$^{-1}$ $\psi(3770)$ data samples, respectively, 
use the Mark III tagging method.   
An alternative CLEO analysis, CLEO(NoTag) \cite{CLEOPolingFPCP}, does not use the $D$-tagging technique 
but ``reconstructs'' the neutrino by fully reconstructing the semileptonic decay from 
all decay products in the event.    
The Belle result \cite{BELLEKpienu} 
uses a 282 fb$^{-1}$ sample collected at or near the $\Upsilon(4S)$ resonance 
with signal events reconstructed from the decay process $D^{*+} \to D^0 \pi^+$.  
For comparison, a recent unquenched LQCD result is also listed in Table \ref{table:DKpienuBF}.  
The experimental results for $D^0 \to K^-(\pi^-) l^+ \nu$ are approaching the precision of 
2$\%$ (4$\%$), while the LQCD predictions are at the $\sim 20\%$ level.

\begin{table}[htb]
\caption{Branching fractions for the decays $D^0 \to K^-e^+ \nu_e$ and $D^0 \to \pi^-e^+ \nu_e$.
Results are given in percent.}
\label{table:DKpienuBF}
\begin{tabular}{lll}
\hline
 & $K^-~e^+ \nu$ & $\pi^-~e^+ \nu$ \\
\hline
PDG04\cite{PDG04}  & 3.58(18)  & 0.36(6) \\
BES\cite{BESKpienu} & 3.82(40)(27)  & 0.33(13)(3) \\
CLEO-c\cite{CLEO56pbD0}$^a$ & 3.44(10)(10)  & 0.262(25)(8) \\
Belle\cite{BELLEKpienu}  & 3.45(10)(19)  & 0.279(27)(16) \\
CLEO(Tag)$^b$  & 3.58(5)(5)  & 0.309(12)(6) \\
CLEO(NoTag)$^b$  & 3.56(3)(11)  & 0.301(11)(10) \\
\hline
LQCD\cite{LQCDSemilepD}  & 3.8(3)(7)  & 0.32(3)(7) \\
\hline
\end{tabular}\\
$^a$From a 56 pb$^{-1}$ sample using the $D$-tag technique and is a subset of the other CLEO results.\\  
$^b$Preliminary results, not to be averaged. 
\end{table}

\begin{table*}[htb]
\caption{Simple pole and modified pole fit results from $D \to K(\pi)l^+ \nu$ decays.}
\label{table:DPlnuPoleFF}
\begin{tabular}{lllll}
\hline
 & $m^{D\to K}_{pole}$ (GeV) & $m^{D\to\pi}_{pole}$ (GeV) & $\alpha^{D\to K}$ & $\alpha^{D\to\pi}$ \\
\hline
FOCUS\cite{FOCUSKpienuFF} & 1.93(5)(3) & 1.91$^{+0.30}_{-0.15}$(7) & 0.28(8)(7) & \\
CLEO III\cite{CLEOIIIKpienuFF} & 1.89(5)$^{+0.04}_{-0.03}$ & 1.86$^{+0.10+0.30}_{-0.06-0.03}$ 
 & 0.36(10)$^{+0.03}_{-0.07}$ & 0.37$^{+0.20}_{-0.31}$(15) \\
BaBar\cite{BABARKenuFF}$^a$ & 1.854(16)(20) &  & 0.43(3)(4) &  \\
Belle\cite{BELLEKpienu} & 1.82(4)(3) & 1.97(8)(4) & 0.52(8)(6) & 0.10(21)(10) \\
CLEO-c(Tag)$^b$ & 1.96(3)(1) & 1.95(4)(2) & 0.22(5)(2) & 0.17(10)(5) \\
CLEO-c(NoTag)$^b$ & 1.97(2)(1) & 1.89(3)(1) & 0.21(4)(3) & 0.32(7)(3) \\
\hline
LQCD\cite{LQCDSemilepD} & 1.72(18) & 1.99(17) & 0.50(4) & 0.44(4) \\
\hline
\end{tabular}\\
$^a$Preliminary results.  $^b$Preliminary results, not to be averaged.  
\end{table*}

The differential partial width for $D \to K(\pi)l^+ \nu$ decays 
is governed by one form factor, $f_+(q^2)$, and is given by
\begin{equation}
\frac{d\Gamma}{dq^2} =\frac{G^2_F}{24\pi^3}~p^3_{K(\pi)}~|V_{cs(d)}|^2~\left|f_+^{K(\pi)}(q^2)\right|^2,
\label{eq:SemilepDecPartWidth}
\end{equation}
where $p_K$ ($p_{\pi}$) is the kaon (pion) momentum in the $D$ rest frame 
and $q^2$ is the invariant mass of the lepton-neutrino pair.  

Various parameterizations of the form factor have been proposed.  
The earliest is in terms of a simple pole model given as 
\begin{equation}
f^{K(\pi)}_+(q^2) = \frac{f^{K(\pi)}_+(0)}{1 - q^2/\left(m^{D\to K(\pi)}_{{\rm pole}}\right)^2}, 
\label{eq:simplepole}
\end{equation}
where the pole mass is expected to be the $D^0 K^+$ ($D^0 \pi^+$) vector state 
$M(D^{*+}_s) = 2.112$ ($M(D^{*+}) = 2.010$) GeV.  
Becirevic and Kaidalov \cite{ModPole} proposed a modified pole model, 
which explicitly incorporates the $D^{*+}_{(s)}$ masses but includes a term $\alpha$ to account 
for deviations from the vector masses.  
Becher and Hill \cite{HillDFF05} proposed a less model-dependent way of dealing with the 
analytic singularities at $q^2 = m^2_{{\rm pole}}$.  
They project the $q^2$ dependence into a different parameter space, 
which pushes the cut singularities far from the physical $q^2$ region, 
and the corresponding form factor can be represented by a rapidly converging Taylor series.  
Hill \cite{HillFPCP06} extended the procedure to explicitly describe semileptonic $D$ decays.

Table \ref{table:DPlnuPoleFF} lists the recent experimental results for the simple and 
modified pole variables.  
The FOCUS result \cite{FOCUSKpienuFF} is determined from $\approx$ 13000 
$D^{*+} \to D^0 \pi^+$ decays, where $D^0 \to K^- (\pi^-)~\mu^+ \nu_{\mu}$. 
The CLEO III \cite{CLEOIIIKpienuFF}, BaBar \cite{BABARKenuFF}, and Belle \cite{BELLEKpienu} results 
studied $D^0$ decays from the decay process $D^{*+} \to D^0 \pi^+$ 
using 7 fb$^{-1}$, 75 fb$^{-1}$, and 282 fb$^{-1}$ data samples, respectively, 
collected at or near the $\Upsilon(4S)$ resonance.  
The preliminary CLEO results (both tagged and untagged analyses) use isospin averaging of 
$D^0 \to K^- e^+ \nu_{e}$  and $D^+ \to K^0_S e^+ \nu_{e}$ decays for $D \to K$ studies and 
$D^0 \to \pi^- e^+ \nu_{e}$  and $D^+ \to \pi^0 e^+ \nu_{e}$ decays for $D \to \pi$ studies.  
The recent unquenched LQCD result \cite{LQCDSemilepD} is also listed in Table \ref{table:DPlnuPoleFF}.  
The measurements of the pole masses as determined using the simple pole model 
are all lower than the expected vector states.  
The results for $\alpha$ are inconsistent with the null result, 
but present experimental accuracy does not constrain it well.  

Table \ref{table:DPlnuf0FF} lists recent results for $f_+^{K(\pi)}(0)$.  
The form factor model for CLEO results use the Hill series parameterization with three parameters, 
while the BES and Belle results use the modified pole model.  
While variations exist between the experimental results, 
they are consistent with the unquenched LQCD result.   

\begin{table}[htb]
\caption{Results for the $f^{K(\pi)}_+(0)$ form factors.  
The form factors are similarly derived using the 2006 global fit values 
for $V_{cs}$ and $V_{cd}$ \cite{PDG06}, with the exception of the Belle result.}
\label{table:DPlnuf0FF}
\begin{tabular}{lll}
\hline
 & $|f_+^{K}(0)|$ & $|f_+^{\pi}(0)|$ \\
\hline
BES\cite{BESKpienu}  & 0.80(4)(3)  & 0.72(14)(3)  \\
Belle\cite{BELLEKpienu}  & 0.695(7)(22)  & 0.624(20)(30)\\
CLEO(Tag)$^a$  & 0.761(10)(7) & 0.660(28)(11) \\
CLEO(NoTag)$^a$  & 0.749(5)(10) & 0.636(17)(13)  \\
\hline
LQCD\cite{LQCDSemilepD} & 0.73(3)(7)  & 0.64(3)(6)  \\
\hline
\end{tabular}\\
$^a$Preliminary results, not to be averaged.
\end{table}

Using the unquenched LQCD prediction for $f^{K(\pi)}_+(0)$ \cite{LQCDSemilepD} 
allows for an experimental measurement of the CKM matrix elements $V_{cd}$ and $V_{cs}$.  
The preliminary $V_{cs}$ result from the CLEO tagged (untagged) analysis is 
$V_{cs} = 1.014\pm 0.013\pm 0.009\pm 0.106$ ($0.996\pm 0.008\pm 0.015\pm 0.104$), 
where the last uncertainty arise from the theoretical uncertainty in $f_+^{K}(0)$.  
The preliminary CLEO result for $V_{cd}$ is $0.234\pm 0.010\pm 0.004\pm 0.024$ 
($0.229\pm 0.007\pm 0.009\pm 0.024$).  
For comparison, the 2006 global fit values are $V_{cs} = 0.97296\pm 0.00024$ 
and $V_{cd} = 0.2271\pm 0.0010$ \cite{PDG06}. 
The experimental uncertainties for $V_{cs}$ and $V_{cd}$ are at the 2$\%$ and 4$\%$ level, 
respectively, while the theoretical uncertainty is on the order of 10$\%$.  
When future theoretical predictions achieve higher precision, 
semileptonic decays will be an ideal environment to precisely measure $V_{cd}$ and $V_{cs}$.

\subsection{$D \to {\rm (vector)}l^+ \nu$}

While the semileptonic $D$ decays to vector mesons are an additional method 
to measure $V_{cs}$ and $V_{cd}$, they are complicated by the presence of three form factors 
associated with the three helicity states of the final state meson.  
The spectroscopic pole dominance model proposes to parametrize the vector mesons in 
terms of a vector and two axial vector form factors, which are defined as 
\begin{equation}
V(q^2) = \frac{V(0)}{1 - q^2/m^2_V},~~
A_i(q^2) = \frac{A_i(0)}{1 - q^2/m^2_{Ai}},
%\label{eq:x}
\end{equation}
where $m_V$ = 2.1 GeV and $m_{A1} = m_{A2}$ = 2.5 GeV.  
The normalized form factors are defined by two ratios, 
the vector to first axial vector $R_V = V(0)/A_1(0)$ 
and the second to first axial vector  $R_2 = A_2(0)/A_1(0)$.

\subsubsection{$D \to \rho~e^+ \nu_e$}

The CLEO Collaboration reported preliminary results for the decay processes 
$D^{+/0} \to \rho^{0/-} e^+ \nu_e \to (\pi^+ \pi^{-/0}) e^+ \nu_e$ 
using the 281 pb$^{-1}$ $\psi(3770)$ data sample and the $D$-tagging technique \cite{CLEOGaoICHEP}.  
They determine the most precise branching fractions of
${\cal B}(D^+ \to \rho^0 e^+ \nu_e) = (2.32\pm0.20\pm0.12)\times10^{-3}$ and 
${\cal B}(D^0 \to \rho^- e^+ \nu_e) = (1.56\pm0.16\pm0.09)\times10^{-3}$.
Using the world average for the $D^{+/0}$ lifetimes \cite{PDG06}, 
they determine the partial width ratio 
$\Gamma(D^0 \to \rho^- e^+ \nu_e)/(2\cdot\Gamma(D^+ \to \rho^0 e^+ \nu_e)) = 0.85\pm0.13$.  
They also performed a simultaneous form factor analysis of the $D^0$ and $D^+$ decays and find 
$R_V = 1.40\pm0.25\pm0.03$ and $R_2 = 0.57\pm0.18\pm0.06$.
This analysis represents the first form factor measurement of a Cabibbo suppressed 
vector decay mode in the charm system.

\subsubsection{$D^+ \to \overline{K}^* l^+ \nu$}

The CLEO collaboration, using the 281 pb$^{-1}$ $\psi(3770)$ data sample 
and the $D$-tagging technique, reported a non-parametric form factor analysis 
using the FOCUS method \cite{FOCUSDpKStarHelFF} for the decay process 
$D^+ \to \overline{K}^* e^+ \nu_{e} \to (K^- \pi^+) e^+ \nu_{\mu}$ \cite{CLEODpKStarHelFF}.  
CLEO confirms the presence of the $s$-wave interference in the $K^- \pi^+$ 
final state and determines results for the form factors to be consist with the FOCUS analysis.  
CLEO did not observe any evidence of $d$- or $f$-wave interference in the $K^- \pi^+$ system.

\section{CONCLUSION}

The recent experimental results for leptonic and semileptonic decays of charm mesons 
are beginning to improve their accuracy to point where theoretical errors are dominating the 
uncertainty in the determination of the $V_{cs}$ and $V_{cd}$ CKM matrix elements.  
These measurements provide stringent tests of theoretical models, which in turn will 
improve the models as to lower their uncertainty but are also used to ``fine-tune'' the models 
so they can be applied with confidence to beauty leptonic and semileptonic decays.  
The experimental precision will continue to improve from the final CLEO-c 
$D$ and $D_s$ data samples and the beginning of data collection with the 
upgraded BES-III and BEPC-II facilities in 2008.

\section{ACKNOWLEDGEMENTS}

I wish to thank Neville Harnew, Guy Wilkinson, and all of the organizers for an 
engaging and enjoyable conference.  This work was supported by the 
U.S. National Science Foundation and Department of Energy.

\end{document}